\documentclass[aps,groupaddress,superscriptaddress,reprint,amsmath,amssymb,graphicx]{revtex4-1}
\usepackage{graphicx}
\usepackage{physics}
\usepackage{bm}
\usepackage{bbold}
\usepackage{xcolor}
\begin{document}
\preprint{APS/123-QED}

\title{Dissipative symmetry breaking in non-equilibrium steady states}

\author{Matteo Sireci}
\email{msireci@onsager.ugr.es}
\affiliation{Departamento de Electromagnetismo y F{\'\i}sica de la Materia and Instituto Carlos I de F{\'\i}sica Te\'orica y Computacional, Universidad de Granada, E-18071 Granada, Spain}
\author{Daniel Maria Busiello}
\email{busiello@pks.mpg.de}
\affiliation{Max Planck Institute for the Physics of Complex Systems, 01187 Dresden, Germany}

\begin{abstract}
\noindent
The connection between dissipation and symmetry breaking is a long-standing enigma in statistical physics.
It is intimately connected to the quest of a non-equilibrium functional whose minimization gives the non-equilibrium steady state (NESS). Writing down such a functional, we show that, in the presence of additive noise, any NESS is characterized by the minimum entropy production compatible with the maximum dissipation along cycles in the trajectory space. This result sheds light on the excess entropy production principle and the onset of chiral symmetry breaking out-of-equilibrium, indicating that the housekeeping dissipation is connected with the tendency of performing cycles in a preferential direction. Finally, when multiplicative noise is present, we find that the non-equilibrium functional has two dissipative symmetry-breaking contributions, one stemming from cycles and the other from a \textit{thermophoresis}-like effect. Our framework paves the way to understand selection phenomena as symmetry-breaking processes driven by non-equilibrium dissipation.
\end{abstract}

\maketitle
How out-of-equilibrium systems self-organize to reach a stable non-equilibrium stationary state (NESS) is an unsolved enigma. Classical general results dictating how NESS should be approached, such as Onsager's minimum dissipation principle \cite{Onsager1,Onsager2} and the Glansdorff-Prigogine criterium \cite{glansdorff1964general}, have attracted recent attention in various contexts \cite{Bertini2004MinimumDP,maes2015revisiting,ito2022information}. However, an agreement on a unified and solid physical picture is still lacking, leaving open the problem of finding first principles valid away from equilibrium. Nevertheless, the last decades witnessed the discovery of fluctuation theorems \cite{gallavotti1995dynamical,jarzynski1997nonequilibrium,kurchan1998fluctuation,crooks1999entropy,lebowitz1999gallavotti}, universal relations holding arbitrarily far  from equilibrium. The core message of these results is that non-equilibrium systems break time-reversal symmetry, preserving a weaker version of it that quantifies dissipation. 

Symmetry breaking mechanisms are also considered at the heart of the emergence of self-organization away from equilibrium, following the inspiration of Prigogine's idea of ``dissipative structures" \cite{prigogine1967symmetry,Anderson1987BrokenSE, solvay}. Although a strict connection was, and is, still debated, recent developments showed that macroscopic systems can spontaneously break some symmetries due to collective effects \cite{Jona_Lasinio_2010, lecomte} or rare large  fluctuations \cite{hurtado2011spontaneous,tizon_order}. From a broader and more fundamental perspective, how dissipation induces a symmetry breaking in the \emph{trajectory} space is a fascinating topic still largely unexplored.

Here, we build a bridge between symmetry breaking and dissipation in mesoscopic non-equilibrium systems, i.e., ruled by Stochastic Thermodynamics. This framework encompasses several well-known experimentally realizable examples, from molecular machines \cite{seifert2012stochastic} to chemical reaction networks \cite{rao2016nonequilibrium}. In recent years, an increasing wealth of studies is investigating the non-equilibrium features of these systems, such as entropy production \cite{van2010three,pigolotti2017generic,busiello2019entropy,busiello2019entropyJSTAT}, current fluctuations \cite{barato2015thermodynamic,falasco2020unifying}, and dissipation-driven asymmetries \cite{astumian2019kinetic,busiello2021dissipation,dass2021equilibrium}. 
Starting from a Langevin dynamics with constant diffusion, we write down a general non-equilibrium functional whose minimization gives the correct stationary state. We show that this is composed of two terms: the first one is the total entropy production and quantifies dissipation, while the second one is minus the housekeeping dissipation. Most importantly, this second contribution is related to broken chiral symmetries in the trajectory space and coincides with the dissipation along cycles (in the simplest case). As a consequence, any stationary state can be understood as the minimum dissipative state compatible with the existence of fluxes along preferential directions, which are absent at equilibrium. Remarkably, this statement corresponds to the minimization of the excess entropy production, which appears to be a more fundamental principle than the standard second law for non-equilibrium systems \cite{van2010three,hatano2001steady,sekimoto1998langevin,ito2022information,dechant2022geometric, ito_non_linear, maes_minimum}. Finally, we unravel the connection between dissipation and chiral symmetry breaking in systems with multiplicative noise. This allows us to derive the most general form of the non-equilibrium functional and generalize the principle of excess entropy production minimization.

Consider an overdamped driven-diffusive system whose evolution is described by the following equation:
\begin{equation}
    \dot{\vec{x}}= \vec{F} + \vec{\xi}(t) \;,
    \label{Langevin}
\end{equation}
where $\vec{x}$ is the  state of the system in dimension $d$, e.g., position in real space, $\vec{F}$ a general force and $\vec{\xi}$ a Gaussian white noise with variance $\langle \xi_{i}(t) \xi_{j}(t') \rangle = 2 \sigma D_{ij} \delta(t-t')$, with $i,j=1,..,d$. The dissipation-fluctuation relation is in general violated, hence the system will eventually reach a non-equilibrium stationary state (NESS) \cite{kubo1966fluctuation}. Here, $\sigma$ controls the amplitude of the noise and, in a thermodynamic context, it can be interpreted as the temperature. First, we consider the diffusion matrix $\hat{D}$ (with elements $D_{ij}$) to be state-independent.
The Fokker-Planck equation (FPE) associated with Eq.~\eqref{Langevin} is:
\begin{equation}
    \partial_t P = - \vec{\nabla}\cdot\left( \vec{F} P - \sigma \hat{D} \vec{\nabla} P \right) = - \vec{\nabla }\cdot\vec{J }\;.
    \label{FPE}
\end{equation}
The stationary solution of the system admits a solenoidal stationary current, i.e., $\vec{\nabla }\cdot\vec{J }_{\rm st}= 0$ (see \cite{mendler2020predicting} for a detailed discussion). Following the pioneering works of Graham and T\'el \cite{graham1984existence,graham1984weak,graham1985weak} and recent developments in Macroscopic Fluctuation Theory \cite{MFT,Garrido_quasi_pot}, the steady-state solution in the weak noise limit reads:
\begin{equation}
    P_{\rm st}(\vec{x}) = \frac{e^{-V(\vec{x})/\sigma}}{Z} \;,
    \label{SolutionWeak}
\end{equation}
where $V(\vec{x})$ is a quasi-potential that can be rigorously defined and estimated via path-integral methods \cite{MFT, supplemental_material}. 
It is worth noting that, in the presence of additive noise, the deterministic fixed point of Eq.~\eqref{Langevin} coincides with the minimum of this potential.
Noticeably, Eq.~\eqref{SolutionWeak} is exact for any linearized stochastic system.
Plugging Eq.~\eqref{SolutionWeak} into the FPE, expanding up to the zeroth order in $\sigma$, and equating terms of the same order, we get two consistency equations:
\begin{equation}
    \vec{v}_{\rm st}\cdot\vec{\nabla} V = 0 \qquad \vec{\nabla}\cdot \vec{v}_{\rm st} = 0 \;,
    \label{topolAdd}
\end{equation}
where $\vec{v}_{\rm st} = \vec{J}_{\rm st}/P_{\rm st}$
is the stationary velocity of the probability current. The first relation states that the $\vec{v}_{\rm st}$ must be tangent to the potential contour lines, while the second one dictates that it also has to be a solenoidal field. An informative way to express these geometrical properties of $\vec{v}_{\rm st}$ in the pedagogical case of a $3D$ system is $\vec{v}_{\rm st} = \vec{\nabla} \times \vec{B}$, i.e., the curl of a field. Most importantly, these features lead to the following decomposition of the force in a solenoidal (\emph{dissipative}) and gradient (\emph{conservative}) part \cite{supplemental_material}:
\begin{equation}\label{force}
 \vec{F}=\vec{\nabla} \times \vec{B}-\hat{D}\vec{\nabla}V \;.
\end{equation}
We will focus on $3D$ systems in this manuscript only for clarity of notation, but the findings of this Letter hold true in any dimension. In fact, later on we will also present a $2D$ Brownian gyrator as an example.
For stochastic systems, the Kullback-Leibler divergence between $P(\vec{x},t)$ and the stationary distribution $P_{\rm st}(\vec{x})$ is known to be a Lyapunov function of the dynamics \cite{schnakenberg1976network}, and is defined as:
\begin{equation}
    D_{KL}(t) = \int d\vec{x} ~P(\vec{x},t) \log \left( \frac{P(\vec{x},t)}{P_{\rm st}(\vec{x})} \right) \;.
\end{equation}
As a consequence, its time-derivative has to be non-positive and vanishes at the steady-state. Thus, the correct stationary solution of Eq.~\eqref{FPE} can be found by minimizing the following functional with respect to $P$ (including also the normalization constraint):
\begin{equation}
    \mathcal{G} = - \frac{dD_{KL}}{dt} = \langle \vec{v} * \vec{v} \rangle - \langle \vec{v} * \vec{\nabla} \times \vec{B} \rangle \;,
    \label{Gfunc}
\end{equation}
that highlights how the geometrical properties of $\vec{v}^*$ enters the game. 
Here, $\vec{\alpha} * \vec{\gamma} = \sigma^{-1} \vec{\alpha}^{T} \hat{D} \vec{\gamma}$ and $\langle \cdot \rangle =\int \cdot P d\vec{x}$.
The first interesting observation is that $\mathcal{G}$ can be written solely in terms of the velocity of probability currents, which appears to be the most natural quantity to characterize out-of-equilibrium dynamics and NESS.

We can go further with the computation, noticing that the first contribution on the r.h.s. of Eq.~\eqref{Gfunc} coincides with the total entropy production, while the second one quantifies the housekeeping heat dissipation \cite{hatano2001steady}:
\begin{equation}
   \dot{S}_{\rm tot}=\langle \vec{v} * \vec{v}\rangle \qquad
    \dot{S}_{\rm hk}=\langle \vec{v} * \vec{\nabla} \times \vec{B} \rangle
    \label{Shk}   
\end{equation}
This equivalence can be obtained starting from the trajectory-dependent formulation. The entropy production along a trajectory in the form of heat is $\dot{s}_{\rm m} = \dot{\vec{x}} * \vec{F}$ using the Stratonovich prescription \cite{sekimoto1998langevin,seifert2012stochastic,supplemental_material}. A part of $\dot{s}_{\rm m}$ is due to the housekeeping heat, i.e., the one necessary to maintain the steady-state distribution, $\dot{s}_{\rm hk} = \dot{\vec{x}} * \vec{v}^{\rm st}$. The average of $\dot{s}_{\rm hk}$ over trajectories readily gives the second term in Eq.~\eqref{Gfunc}. As a result, the housekeeping dissipation emerges from the non-conservative part of the force and, as we will show later, is linked to the tendency of performing cycles in the trajectory space. 
Hence, in analogy to \cite{van2010three}, we have:
\begin{equation}
  \mathcal{G} = \dot{S}_{\rm tot} -\dot{S}_{\rm hk} = \dot{S}_{\rm ex} \geq 0 \;.
  \label{excess}
\end{equation}
The correct steady state of a general non-equilibrium dynamics is given by the minimum excess entropy production. Eqs.~\eqref{Gfunc} and \eqref{Shk} give a geometrical meaning to this principle and constitute the first result of this Letter.


This minimum $\dot{S}_{\rm ex}$ criterion resembles the Glandsdorff-Prigogine principle and trivially corresponds to the minimum entropy production in the equilibrium case \cite{prigogine1968introduction,maes_minimum, maes2015revisiting}. In our framework, we can improve the physical interpretation of this result. The first term of $\mathcal{G}$ in Eq.~\eqref{Gfunc} only dictates that the system tends to minimize its total dissipation, as for equilibrium relaxation phenomena. The second contribution emerges only out of equilibrium, hence encoding an additional dissipation stemming from the properties of the steady-state velocity. Indeed, Eq.~\eqref{Gfunc} explicitly shows that this extra dissipative term manifests into a symmetry breaking in the trajectory space, since the velocity tends to maximize the dissipation along preferential directions. In other words, the NESS is the least dissipative state compatible with a velocity that is maximally aligned with the closed force lines of $B$, thus accounting for a tendency to circulate and a consequent stationary dissipation into the environment.

To better characterize the connection between dissipation and emergent symmetry breaking in NESS, we study the probability to observe a closed trajectory. The Onsager-Machlup action \cite{cugliandolo2017rules} using the Stratonovich prescription for the system in Eq.~\eqref{Langevin} is:
\begin{equation}
    \mathcal{S}(\Gamma) = \frac{1}{2}\int_0^t d\tau \left[ \frac{1}{2} \left( \dot{\vec{x}} - \vec{F} \right)^{T} \hat{D}^{-1} \left(  \dot{\vec{x}} - \vec{F} \right) - \vec{\nabla}\cdot\vec{F} \right] \;,
\end{equation}
where $\Gamma$ is a trajectory of duration $t$, from $\vec{x}_{0}$  to $\vec{x}_{f}$ along which all quantities have to be evaluated. The asymmetric part of the action, $\mathcal{S}^{\rm a}$, is related to the dissipation along a trajectory, which is also equal to the ratio between the probability of $\Gamma$ and its time-reversal, $\tilde{\Gamma}$ \cite{seifert2012stochastic}. By considering any closed trajectory $\Gamma_\circ$, i.e., where $\vec{x}_{0}=\vec{x}_{f}$, the terms accounting for the initial and final states of the trajectory vanish, thus we have:
\begin{equation}
    \mathcal{S}^{\rm a} = \log\frac{P(\Gamma_\circ)}{\tilde{P}(\tilde{\Gamma}_\circ)} = \int_0^t d\tau ~\dot{\vec{x}} * \vec{F} = \int_{\Gamma_\circ} d\vec{x} * \vec{\nabla} \times \vec{B} \;.
\end{equation}
This establishes a clear connection between the propensity of performing closed trajectories in a preferential direction and the housekeeping dissipation. Indeed, $\dot{S}_{\rm hk}$ enters into the non-equilibrium functional counterbalancing the entropy production minimization, Eq.~\eqref{excess}, hence breaking the chiral symmetry (i.e., clockwise or counter-clockwise) that is present at equilibrium. This finding clarifies the main result of this Letter, providing also a broader context for recent results about a topological fluctuation theorem \cite{mahault2022topological} and gauge symmetries in thermodynamics \cite{Polettini_2012,polettini_dice,wang_gauge}.

The proposed framework extends beyond the case of additive noise to multiplicative noise scenarios, where the Glansdorff-Prigogine principle has not been formulated. Notably, a state-dependent diffusion coefficient emerges in systems affected by thermal gradients \cite{busiello2021dissipation, dass2021equilibrium} or fluctuating environments \cite{chechkin2017brownian,nicoletti2021mutual}. Moreover, finite-size fluctuations are important in fluctuating hydrodynamics \cite{Garrido_2021} and field theories \cite{te2020classical,cornalba2019regularized}.
Consider the following generic Langevin equation :
\begin{equation}
    \dot{\vec{x}} = \vec{F}+\hat{G}(\vec{x}) \vec{\xi}(t)
    \label{LangevinMult} \;,
\end{equation}
with $\vec{\xi}$ Gaussian white noises with correlation matrix $\hat{C}$, $\hat{G}$ the state-dependent part of the diffusion matrix that we consider diagonal for simplicity, and hence a total diffusion matrix $\hat{D}=\hat{G}^{T}\hat{C}\hat{G}$. For $\hat{G} = \hat{1}$ (the identity matrix), we go back to Eq.~\eqref{Langevin}. This choice allows us to write the following stationary solution of Eq.~\eqref{LangevinMult} in the weak-noise limit:
\begin{equation}
    P_{\rm st}(\vec{x}) = z(\vec{x}) \eta(\sigma) \frac{e^{-V(\vec{x})/\sigma}}{Z} \;.
    \label{MultSol}
\end{equation}
Here $z$ and $\eta$ are functions of space and noise amplitude, respectively, whose form depends on the model. The first important observation is that the maximum of the probability distribution does not coincide with the deterministic fixed point, because of the presence of a space-dependent noise. 
Moreover, the stationary velocity now contains the sum of a zeroth and a first order term in the noise amplitude: $\vec{v}_{\rm st} = \vec{v}_{\rm st}^{(0)} + \sigma \vec{v}_{\rm st}^{(1)}$. The zeroth order velocity is tangent to the height lines of $V$, as in the additive case. We show that this condition actually implies that the $\vec{v}_{\rm st}^{(0)}$ is, again, a solenoidal field, i.e. $\vec{v}_{\rm st}^{(0)}=\vec{\nabla}\times\vec{B}$ \cite{supplemental_material}. Furthermore, the geometrical properties of the first order velocity can be unveiled by using a specific change of variables. Here, we consider Stratonovich integration, even if our result can be derived for any prescription, as we show in \cite{supplemental_material}. We remind that a change of variables would allow to map a multiplicative noise into an additive one, restoring the results we obtained before, but in the transformed space. In particular, choosing the Jacobian of the transformation $\vec{x} \to \vec{x}'$ as $\hat{\Lambda} = \hat{G}^{-1}$, the diffusion matrix of the transformed dynamics is equal to $\hat{C}$. By deriving how the probability distribution and the velocity transform under this change of variables, we determine that $z(\vec{x})=\vert \hat{\Lambda}\vert=\vert \hat{G}\vert^{-1}$, where $|\cdot|$ indicates the determinant. As a consequence, $\vec{v}_{\rm st}^{(1)}=\hat{D}\vec{\nabla}\psi$, where $\psi=\log\vert \hat{G}\vert$, i.e., a gradient field that contributes with a new term in the functional \cite{supplemental_material}:
\begin{gather}
    \mathcal{G} = \langle \vec{v} * \vec{v} \rangle - \langle \vec{v} * \vec{\nabla} \times \vec{B} \rangle -\sigma\langle \vec{v} * \hat{D}\vec{\nabla}\psi \rangle.
    \label{GfuncMult}
\end{gather}
In analogy to Eq.~\eqref{Gfunc}, the first contribution quantifies the total entropy production, which tends to be minimized as the system goes toward stationarity. The second and third terms amount to the dissipated heat to maintain the steady-state and depends on the symmetries that are broken in the trajectory space. These terms counterbalance the entropy production minimization and play a role analogous to the housekeeping heat in the additive noise case. In particular, the second term accounts for the heat dissipated along the solenoidal part of the force, while the third contribution is proportional to the derivative of $|G|$ and vanishes in the limit of additive noise. An intuitive understanding of this last term might come considering a system subjected to a thermal gradient, $T(\vec{x})$. In this case, it can be readily shown that $\vec{\nabla}\psi \propto T^{-1} \vec{\nabla} T$, resembling an additional dissipation arising from thermophoretic effects and due to the necessity of transporting heat \cite{liang2022emergent}. A general physical interpretation of $\psi$ might be particularly challenging to find, since the multiplicative noise arises in a wide variety of systems, and might be the topic of future investigations.

Eq.~\eqref{GfuncMult} constitutes the second main result of this Letter. In this more complex scenario, the non-equilibrium functional $\mathcal{G}$ does not coincide with the excess entropy production, and its thermodynamic properties are not known a priori. We find that it can always be interpreted as the sum between the excess entropy production 
and an additional \textit{thermophoretic} term, named after its meaning in thermally-driven systems and indicating the accumulation of probability in regions of low noise. This result generalizes the Glandsdorff-Prigogine principle to generic (linear and non-linear) stochastic systems away from equilibrium.


Here, we present two pedagogical examples where the chiral symmetry breaking and the emergence of preferential cyclic trajectories accountable for the housekeeping entropy production are easy to  visualize.
Consider a $3D$ Brownian particle confined on a torus by a (quadratic) potential, and driven along the torus itself by a constant non-conservative force $f$ which induces a clear chiral symmetry breaking.
It is instructive to write the system in polar coordinates $(\rho,\phi,z)$, as the stationary solution is found to support a probability flux only along the angle variable, $\phi$. 
Hence, by evaluating the steady-state flux and dividing it by the steady-state distribution, we obtain a stationary velocity $\vec{v}_{\rm st} = - (f/\rho) \hat{v}_\phi$, where $\hat{v}_\phi$ is the versor associated with $\phi$. 
As expected, $\vec{v}_{\rm st}$ is a solenoidal field and indicates that the housekeeping heat stems from cyclic trajectories running across the entire torus in a preferential direction \cite{supplemental_material}. In this example, the chiral symmetry breaking is explicitly linked to the symmetry of the non-conservative driving, making immediate to identify the origin of the term in $\mathcal{G}$ that tends to be maximized in the NESS. 

\begin{figure}[t]
    \centering
    \includegraphics[width=0.9\columnwidth]{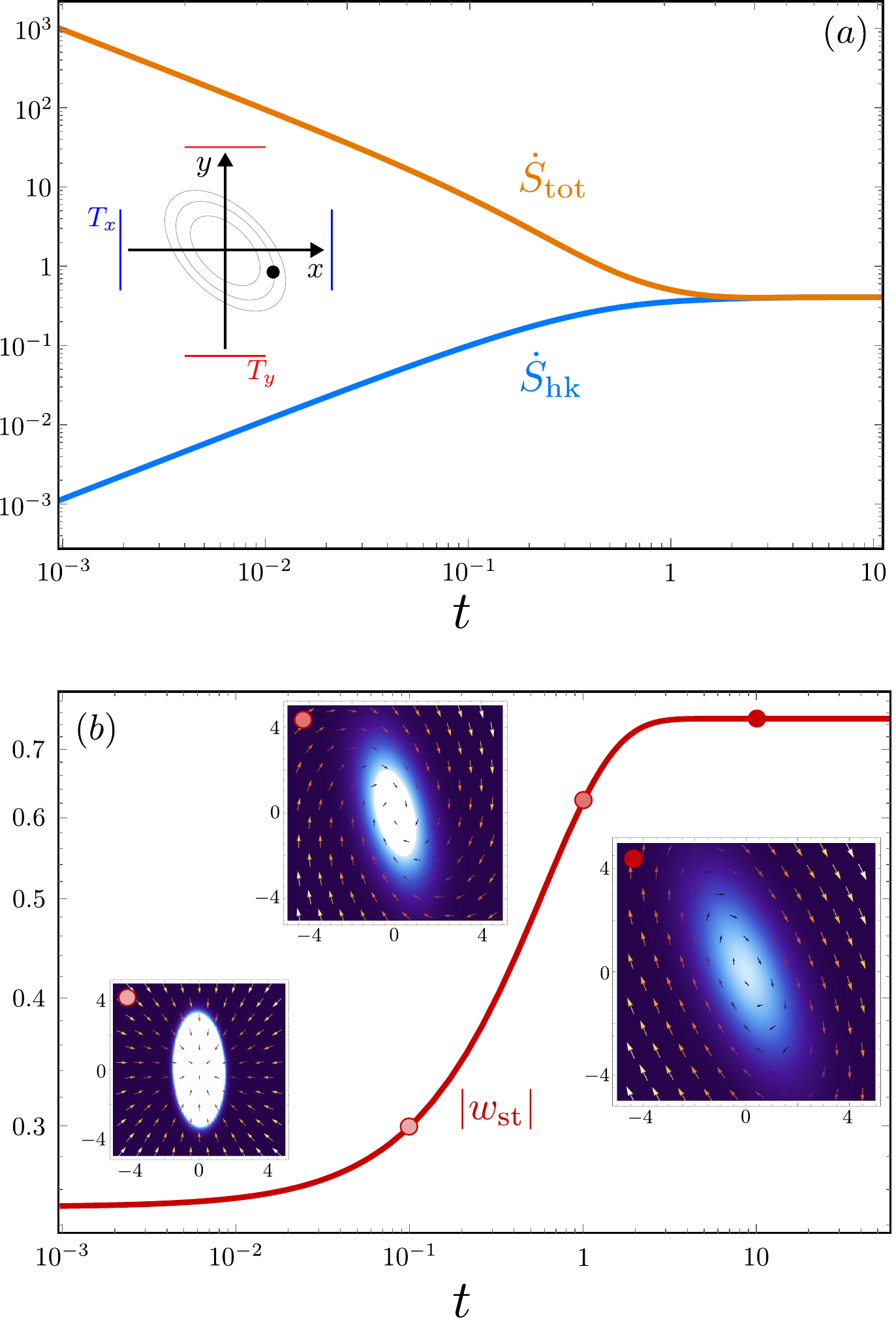}
    \caption{Dissipative symmetry-breaking in a Brownian gyrator. (a) A sketch of the model is presented. As a function of time, we show that the system tends to minimize $\dot{S}_{\rm tot}$, while maximizing $\dot{S}_{\rm hk}$, which is the dissipation associated with the symmetry breaking in the trajectory space. At the NESS, $\dot{S}_{\rm tot} = \dot{S}_{\rm hk}$. (b) The modulus of the vorticity is plotted. The inset show the pdf (in color-scale) and the velocity vector field (bigger arrows corresponds to a stronger field) as time increases. Insets clearly indicate that the tendency to rotate in a preferential direction increases towards the NESS.}
    \label{fig2}
\end{figure}

The second example is a $2D$ Brownian gyrator, i.e., a diffusive particle in a confining potential subjected to two reservoirs at different temperatures, $T_x$ and $T_y$, each one acting along one direction. 
The Langevin equation describing the system is:
\begin{eqnarray}
    \dot{x} &=& - \partial_{x}U(x,y)+ \sqrt{2 T_x} \xi_x(t), \\
    \dot{y} &=& - \partial_{x}U(x,y) +\sqrt{2 T_y} \xi_y(t) \nonumber ,
\end{eqnarray}
where $U(x,y) = (1/2)(x^2 + y^2) + uxy$, with $u$ the asymmetry of the elliptic potential ($|u|<1$ to ensure stability), and $\xi$ a Gaussian white noise with unit variance. Since $T_x \neq T_y$ but the mobilities along $x$ and $y$ are the same, the fluctuation-dissipation relation cannot be satisfied and the system breaks detailed balance \cite{kubo1966fluctuation}. The stationary distribution can be explicitly obtained in the form of the weak-noise ansatz, Eq.~\eqref{SolutionWeak} \cite{cerasoli2018asymmetry}. Thus, we can find the steady-state velocity, $\vec{v}_{\rm st} = \vec{\nabla} U - \vec{T}/\sigma \cdot \vec{\nabla} V$, where $V$ and $\sigma$ are functions of $u$, $T_x$, and $T_y$ and $\vec{T} = (T_x, T_y)$. It is readily verified that the gradient of $\vec{v}_{\rm st}$ is zero. Since we are studying a $2D$ system, it is instructive to quantify the onset of the chiral symmetry breaking by computing the vorticity, $w_{\rm st} = \partial_x (\vec{v}_{\rm st})_y - \partial_y (\vec{v}_{\rm st})_x$, i.e., the $2D$-curl of the velocity. In fact, $w_{\rm st}$ is connected to the tendency of performing closed trajecories and it reads \cite{supplemental_material}:
\begin{equation}
    w_{\rm st} = \frac{4 T_x u}{4 T_x T_y + (T_x - T_y)^2 u^2} \left( T_x^2 - T_y^2 \right)
\end{equation}
Interestingly, the chiral symmetry breaking comes from the simultaneous presence of a temperature difference and an interaction coefficient, $u$. Since this model can be solved analytically for all times, in Fig.~\ref{fig2}a we show the temporal evolution of $\dot{S}_{\rm tot}$ and $\dot{S}_{\rm hk}$, highlighting that the steady state coincides with a minimization of the total entropy production compatible with the maximum dissipation due to chiral symmetry breaking. While converging to stationarity, the system selects a region of the space where to concentrate probability fluxes, resulting in a stationary vorticity (see Fig.~\ref{fig2}b).

Our approach sheds light on the role of topological aspects in NESS. Indeed, the housekeeping dissipation is associated with the onset of vortex structures around the zero-current point that acts as an \textit{emerging} topological defect (see Fig.~\ref{fig2}b). Vortices arise around the deterministic fixed point only in the presence of additive noise, otherwise complex structures (e.g., dipole currents) might emerge \cite{mendler2020predicting}. Moreover, beyond the simple $2D$ scenario, zero-current manifolds might appear and act as defects, hence generating non-trivial vortex structures at NESS and enriching even more the presented picture \cite{mahault2022topological}.


The idea that some kind of selection can naturally take place out-of-equilibrium due to dissipation-driven processes is gaining momentum in the field of physical chemistry \cite{busiello2021dissipation,dass2021equilibrium} and evolutionary dynamics \cite{PRICE}. Our framework might provide a useful tool to tackle this problem. At stationarity, probability fluxes will be focused in determined regions of the variable space, $\vec{x}$. If each $x_{i}$ represents a species in a fictitious space, the emergent chiral symmetry breaking can reflect the onset of preferential cycles involving some of them. Additionally, these cycles are intimately connected with the housekeeping dissipation and could be explored using large-deviation theory \cite{cycle_random_walk}. This perspective intriguingly resembles the idea of \textit{hypercycles} in the context of the origin of life \cite{eigen1978hypercycle}. In order to push forward this analogy, we leave for future investigation the extension of the framework discrete-state dynamics (e.g., master equations), in line with preliminary works in this direction \cite{ito2022information, ito_non_linear}.

Active matter represents another possible field of application of the presented framework. In the simplest case of an active Brownian particle \cite{dabelow2019irreversibility}, for example, a solution can be readily found at all times and it closely resembles the one of the $2D$ Brownian gyrator. The conceptual leap, in this context, is that emergent dissipative cycles arise in the space describing the particle and the bath, making the physical interpretation harder \cite{active_engine}. Future works might also explore this direction of research.

To summarize, we unraveled the connection between dissipation and chiral symmetry breaking in NESS of stochastic mesoscopic systems. We showed that the housekeeping dissipation is intimately connected to the tendency of performing cycles in the trajectory space. This quantity tends to be maximized in the NESS and counterbalances the entropy production minimization. In the most general scenario of multiplicative noise, the non-equilibrium functional shows two dissipative symmetry-breaking contributions. The first accounts for cycles in a transformed space, while the second can be seen as a thermophoretic dissipation. This results generalizes the principle of excess entropy production minimization, extends and clarifies the physical meaning of the Glansdorff-Prigogine principle, and paves the way to understand selection phenomena in different contexts as a result of a symmetry breaking process driven by non-equilibrium dissipation.

\acknowledgments
M.S. acknowledges G. Jona-Lasinio for early inspiration in studying non equilibrium statistical physics, M.A. Muñoz, P. Garrido and P.I. Hurtado for teaching the bases of MFT and of non-eq. potentials. D.M.B. thanks M. Ciarchi (Retoño) for illuminating discussions on symmetry breaking.
We are grateful to L. Peliti for a detailed reading of the manuscript and many comments, and to E. Roldan, M.A. Muñoz, P.I. Hurtado, C. Espigares, R. Hurtado and C. Maes for important discussions and insights. M.S. acknowledge the Spanish Ministry and Agencia Estatal de investigaci{\'o}n (AEI) through Project of I+D+i Ref. PID2020-113681GB-I00, financed by MICIN/AEI/10.13039/501100011033 and FEDER "A way to make Europe", as well as the Consejer{\'\i}a de Conocimiento, Investigaci{\'o}n Universidad, Junta de Andaluc{\'\i}a and European Regional Development Fund, Project P20-00173 for financial support.

\clearpage
\newpage

\section*{Supplemental Material}
Here, we present the derivation of all the results presented in the main text, with additional mathematical steps and considerations. We aim to create a self-consistent (and more detailed) file as supplemental material, in which we relate geometrical properties of the velocity with dissipative features.

\section{Constant diffusion.}
\subsection{Geometrical and topological properties of current velocities}
Consider an overdamped driven-diffusive system whose evolution is described by the following equation (in the covariant formulation):
\begin{equation}
    \dot{x}^\mu = F^\mu + \xi^\mu(t) \;,
    \label{Langevin}
\end{equation}
where $\vec{x}=(x^{\mu})$, $\mu=1,\dots,d$, is the state of the system, e.g., position in real space, $\vec{F}$ a general force and
$\xi^\mu$ a Gaussian white noise with variance $\langle \xi^\mu(t) \xi^\nu(t') \rangle = 2\sigma D^{\mu\nu} \delta(t-t')$. The dissipation-fluctuation relation is in general violated, hence the system will eventually reach a non-equilibrium stationary state (NESS) \cite{kubo1966fluctuation}. Here, $\sigma$ controls the amplitude of the noise and, in a thermodynamic context, it can be interpreted as the temperature. For simplicity, we consider the diffusion matrix to be state-independent. Later on, we will generalize our framework to multiplicative noise. Let us remind to the reader that in the covariant formalism \cite{Graham1977CovariantFO} any vector, or matrix, element is represented with a up greek index, i.e $x^{\mu}$, $D^{\mu\nu}$ or $\partial^{\mu}=\frac{\partial}{\partial x_{\mu}}$.
Analogously, the element of the inverse of a matrix is represented with down indices: $(\hat{D}^{-1})^{\mu\nu}=D_{\mu\nu}$.
Any index repetition , necessarily one up and one down, means a sum over such index; for example the scalar product between two vectors $\vec{\alpha}$ and $\vec{\beta}$ reads:
\begin{eqnarray}
    \vec{\alpha}\cdot\vec{\beta}=\alpha^{\mu}\beta_{\mu}=\sum^{d}_{\mu=1}\alpha^{\mu}\beta^{\mu}.
\end{eqnarray}
Note that in this case the matrix for the scalar product is the trivial flat one.
The Fokker-Planck equation (FPE) associated with Eq.~\eqref{Langevin} is:
\begin{equation}
    \partial_t P = - \partial_\mu \left( F^\mu P - \sigma D^{\mu\nu} \partial_\nu P \right) = - \partial_\mu J^\mu \;.
    \label{FPE}
\end{equation}
The stationary solution of the system admits a solenoidal stationary current, i.e., 
\begin{eqnarray}\label{eq:stationariety}
   \partial_\mu J_{\rm st}^\mu = 0 
\end{eqnarray}
(see \cite{mendler2020predicting} for a detailed discussion). Following the pioneering works of Graham and T\'el \cite{graham1984existence,graham1984weak,graham1985weak} and recent developments in Macroscopic Fluctuation Theory \cite{MFT,Garrido_quasi_pot}, the steady-state solution in the weak noise limit reads:
\begin{equation}\label{eq:potential}
    P_{\rm st}(\vec{x}) = \frac{e^{-V(\vec{x})/\sigma}}{Z} \;,
\end{equation}
where $V(\vec{x})$ is a non-equilibrium potential, or quasi-potential. Despite its analytical intractability, there are numerous numerical and empirical methods to estimate $V(\vec{x})$,
where the potential is exactly defined as the minimum of the action along a trajectory to create a fluctuation $\vec{x}$ \cite{MFT}:
\begin{eqnarray}
V(\vec{x})&=&V(\vec{x}^*)+\\&+&\min\left(\frac{1}{2}\int^{0}_{-\infty}dt\left[\frac{1}{2}(\dot{\vec{x}}-\vec{F})\hat{D}^{-1}(\dot{\vec{x}}-\vec{F})-\vec{\nabla}\cdot\vec{F}\right]\right)\nonumber
\end{eqnarray}
where $\vec{x}^*$ is the minimum of the potential, corresponding to the deterministic fixed point. Using the introduced relations, it is possible to decompose the thermodynamic force $\vec{F}$ into a conservative contribution given by the gradient of the quasi-potential and a dissipative one proportional to the stationary current:
\begin{eqnarray}
    J^{\mu}_{\rm st}&=&F^{\mu}P_{\rm st}-\sigma D^{\mu\nu}\partial_{\nu}P_{\rm st}\\
    F^{\mu}&=&J^{\mu}_{\rm st}P^{-1}-D^{\mu\nu}\partial_{\nu}V=v^{\mu}_{\rm st}-D^{\mu\nu}\partial_{\nu}V,
\end{eqnarray}
where we have defined the current velocity as:
\begin{eqnarray}
    v^{\mu}=\frac{J^{\mu}}{P}.
\end{eqnarray}
Given that the quasi-potential contributes with a gradient, and the stationary current is a solenoidal field \cite{mendler2020predicting}, one would be tempted to assume that also the velocity is solenoidal, obtaining in this way a complete geometrical decomposition of the force. Such an assumption is not always true, as it can be seen by recasting the stationary condition, Eq.~\eqref{eq:stationariety} in terms of the velocity:
\begin{eqnarray}
  \nabla\cdot v^*=\frac{1}{\sigma}v^*\cdot \nabla V  
\end{eqnarray}
Hence, the divergence of $\vec{v}_{\rm st}$, i.e. its propensity to be a sink, is just the scalar product between the velocity and the gradient of the potential. To uncover the geometric properties of the system, we employ the quasi-potential ansatz, Eq.~\eqref{eq:potential}, to solve the Fokker-Planck Equation in perturbative orders of $\sigma$ \cite{graham1984weak}:
\begin{eqnarray}\label{eq:stationary}
\partial_{t}P_{\rm st}&=&
\sigma^{-1}P_{\rm st}\left(F^{\mu}+D^{\mu\nu}\partial_{\nu}V\right)\partial_{\mu} V+\nonumber\\&-&\left(\partial_{\mu}F^{\mu}+D^{\mu\nu}\partial_{\mu}\partial_{\nu}V\right)P_{\rm st} =0.
\end{eqnarray}
The first contribution is of order $1/\sigma$:
\begin{eqnarray}
(F^{\mu}+D^{\mu\nu}\partial_{\nu}V)\partial_{\mu}V=0
\end{eqnarray}
Graham interprets this equation as a Hamilton-Jacobi equation with position $q^{\mu}=x^{\mu}$ and moment $p^{\mu}=\partial^{\mu}V$, and can be ideally solved with the method of characteristics. Furthermore, inspired by the work of Jona-Lasinio and collaborators \cite{clausius}, we interpret it as a perpendicularity condition for the stationary current velocity:
\begin{eqnarray}\label{eq:perp_v}
\vec{v}_{\rm st}\cdot\vec{\nabla} V=0.
\end{eqnarray}
This relation tells us that the stationary velocity is always perpendicular to the gradient of the potential, i.e. it is tangent to the potential height lines. As a consequence, as already observed in \cite{mendler2020predicting}, the fixed point of the velocity must coincide with the potential minimum.

Then, the second (and last) contribution is of order $\sigma^0$:
\begin{eqnarray}
\partial_{\mu}F^{\mu}+D^{\mu\nu}\partial_{\mu}\partial_{\nu}V=0,
\end{eqnarray}
and naturally imposes that the stationary velocity is also a divergence free field, i.e. a solenoidal field:
\begin{eqnarray}\label{eq:div_v}
\vec{\nabla}\cdot \vec{v}_{\rm st}=0,
\end{eqnarray}
such that in three dimension it can be written, like the current, as the the curl of a vector field $\vec{B}$:
\begin{eqnarray}
\vec{v}_{\rm st}=\vec{\nabla}\times \vec{B}.
\label{vstar}
\end{eqnarray}
In particular, note that the vorticity of the stationary velocity can be easily calculated:
\begin{eqnarray}
\vec{w}_{\rm st}=\vec{\nabla}\times \vec{v}_{\rm st}=\vec{\nabla}\times \vec{F}+\vec{\nabla}\times(D\vec{\nabla} V),
\end{eqnarray}
where the second contribution is non-zero for anisotropic diffusion.
In the case of a diagonal diffusion matrix in two dimension, one obtains:
\begin{eqnarray}
\vec{w}_{\rm st}&=\vec{\nabla}\times \vec{F}+(D_{xx}-D_{yy})\partial_{x}\partial_{y}V,
\end{eqnarray}
indicating that the vorticity depends on the curl of the force and on the asymmetry of the diffusion coefficients in the two directions. Finally, decomposing the velocity at any time in its stationary and relaxation part, we obtain:
\begin{eqnarray}
    v^{\mu}(t)&=&v^{\mu}_{\rm st}+\tilde{v}^{\mu}\\
    \tilde{v}^{\mu}&=&-\sigma D^{\mu\nu}\partial_{\nu}\log\frac{P(x,t)}{P_{\rm st}}=-\sigma D^{\mu\nu}\partial_{\nu}\phi(x,t)
\end{eqnarray}
where we defined:
\begin{eqnarray}
\phi(x,t)=\log\frac{P}{P_{\rm st}}=\log P(x,t)+\frac{V(x)}{\sigma}.   
\end{eqnarray}
Hence, the relaxation part $\tilde{v}$ is geometrically a gradient and goes to zero at stationarity. The result presented here clarifies a long-lasting comparison between the geometrical properties of non-equilibrium thermodynamics and electromagnetism. Indeed, equilibrium thermodynamics present only a gradient field, like the electric one, while non-equilibrium conditions add a solenoidal field, analogous to the the magnetic one. References to this analogy are scattered across the literature: for example see the introduction to non-equilibrium statistical physics by T.Chou, K.Mallick and R. Zia \cite{Chou_2011}, the works on non-equilibirum landscape by J. Wang and collaborators \cite{wang_rev} and various works in MFT \cite{Bertini2004MinimumDP}.

\subsection{Thermodynamics}
Here, we show that various thermodynamic quantities are naturally expressed in terms of the velocities and have an interesting geometrical meaning.

First of all, the change in the system entropy can be rewritten using geometrical information:
\begin{equation}
\label{eq:entropy}
\dot{S}_{sys}(t)=-\int dx \partial_{t}P(t)\log P(t)=\dot{S}_{tot}(t)-\dot{S}_{env}(t),
\end{equation}
where the total entropy production is
\begin{eqnarray}
\dot{S}_{tot}(t)=\int dx \frac{J^{\mu}D_{\mu\nu}J^{\nu}}{\sigma P}=\langle \vec{v}* \vec{v}\rangle \geq 0,
\end{eqnarray}
and the entropy flux to the environment is
\begin{eqnarray}
\dot{S}_{env}(t)= \langle \vec{v} * \vec{F}\rangle,
\end{eqnarray}
with the diffusive scalar product indicated by the asterisk in which the diffusion matrix acts as a metric \cite{Graham1977CovariantFO}, i.e.:
\begin{eqnarray}
\vec{\alpha} * \vec{\beta}=\alpha^{\mu}D_{\mu\nu}\beta^{\nu};
\end{eqnarray}
and
\begin{eqnarray}
\vec{\alpha} * \vec{\alpha}=\vert \vec{\alpha}\vert^2_{D}.
\end{eqnarray}
The average is performed  as follows:
\begin{eqnarray}
\langle O\rangle =\int O(x) dxP(x,t).
\end{eqnarray}
Thus, the total entropy production can be decomposed in two non-negative parts (the so-called adiabatic decomposition \cite{Glansdorff1971ThermodynamicTO,maes2015revisiting,van2010three}):
\begin{eqnarray}
\dot{S}_{tot}&=&\langle \vec{v}* \vec{v}\rangle = \langle(\vec{\tilde{v}}+\vec{v}_st)*(\vec{\tilde{v}}+\vec{v}_st)\rangle=\nonumber\\ &=&\langle \vec{v}_{\rm st}*\vec{v}_{\rm st}\rangle+\langle\vec{\tilde{v}}*\vec{\tilde{v}}\rangle\nonumber\\
&+&\langle\vec{v}_{\rm st}*\vec{\tilde{v}}\rangle+ \langle \vec{\tilde{v}}* \vec{v}_{\rm st}\rangle,
\end{eqnarray}
where the last line is zero because
\begin{eqnarray}\label{eq:excess_perp}
\langle\vec{v}_{\rm st}*\vec{\tilde{v}}\rangle &=&-\frac{\sigma}{2}\int dx P(x,t) v^{\mu}_{\rm st}D_{\mu\nu}D^{\nu\lambda}\partial_{\lambda}\phi(x,t)\nonumber\\
&=&-\frac{\sigma}{2}\int dx P(x,t) v^{\mu}_{\rm st}\partial_{\mu}\phi(x,t)\nonumber\\
&=&-\frac{\sigma}{2}\int dx \frac{P(t)J^{\mu}_{\rm st}}{P_{\rm st}}\left(\frac{\partial_{\mu}P}{P}+\frac{\partial_{\mu}V}{\sigma}\right)\nonumber\\
&=&\frac{\sigma}{2}\langle \vec{\nabla}\cdot \vec{v}_{\rm st}\rangle-\frac{1}{2}\langle \vec{v}_{set}\cdot \vec{\nabla}V\rangle =0,
\end{eqnarray}
for the perpedicularity conditions (Eq.s~\eqref{eq:perp_v} and \eqref{eq:div_v}). Hence, we obtain the following expression:
\begin{eqnarray}
\dot{S}_{tot}&=&\langle \vec{v}* \vec{v}\rangle=\langle \vert \vec{v}\vert^2_{D}\rangle=
\langle \vec{v}_{\rm st}* \vec{v}_{\rm st}\rangle+\langle\vec{\tilde{v}}*\vec{\tilde{v}}\rangle\nonumber\\
&=& \langle \vert \vec{v}_{\rm st}\vert^2_{D}\rangle+\langle \vert \vec{\tilde{v}}\vert^2_{D}\rangle=
\langle \vert \vec{\nabla} \times \vec{B}\vert^2_{D}\rangle+\sigma^2\langle \vert \vec{\nabla} \phi\vert^2_{D}\rangle.\nonumber\\
\end{eqnarray}
The two contributions corresponds to housekeeping (or adiabatic) and excess (or non-adiabatic) components \cite{van2010three}
\begin{eqnarray}\label{eq:decomp_excess}
\dot{S}_{tot}(t)&=&\dot{S}_{hk}(t)+\dot{S}_{ex}(t)\\
\dot{S}_{hk}(t)&=&\langle\vec{v}* \vec{v}_{\rm st}\rangle =\langle\vec{v}_{\rm st}* \vec{v}_{\rm st}\rangle= \dot{S}_{a}(t)\nonumber\\
&=&\langle \vert \vec{\nabla} \times \vec{B}\vert^2_{D}\rangle\geq 0\\
\dot{S}_{na}(t)&=& \langle\vec{\tilde{v}}*\vec{\tilde{v}}\rangle = \sigma^2\langle \vert \vec{\nabla} \phi(x,t)\vert^2_{D}\rangle\nonumber\\
&=&\dot{S}_{ex}(t) \geq 0
\end{eqnarray}
where in the first line we have used the perpendicularity condition, Eq.~\eqref{eq:excess_perp}. Notice that the housekeeping part is zero only at equilibrium and corresponds to the stationary entropy production. Conversely, the excess part is zero at any NESS and corresponds to the definition given by Prigogine (see \cite{Glansdorff1971ThermodynamicTO,maes2015revisiting}).
Finally, let us note that the time derivative of the average potential gives the balance between the house-keeping entropy production and the entropy flux to the environment, i.e.,
\begin{eqnarray}\label{eq:evolution_potenial}
\frac{d_{t} \langle V\rangle}{\sigma}&=&\langle \vec{v}(t)\cdot \vec{\nabla} V\rangle =\langle \vec{v}* \vec{v}_{\rm st}\rangle -\langle \vec{v}* \vec{F}\rangle\\
&=&\dot{S}_{hk}-\dot{S}_{env},
\end{eqnarray}
and hence must be zero at stationarity. 

\subsection{Lyapunov functional}
Consider the following functional:
\begin{eqnarray}
D_{KL}(t)=\int d\vec{x} P(\vec{x},t)\log\left(\frac{P(\vec{x},t)}{P_{\rm st}(\vec{x})}\right)
\end{eqnarray}
i.e, the Kullback-Leibler divergence between the probability density  function at time $t$ and the stationary one. $D_{KL}$ is known to be a Lyapunov function of the dynamics \cite{schnakenberg1976network}. Note that, in thermodynamic context, $\sigma = k_{B} T$, so that such a functional is akin to a dynamic free energy:
\begin{eqnarray}
    \mathcal{F} = k_{B}T D_{KL}(t) = \langle V \rangle - k_{B}T S(t).
\end{eqnarray}
The time-derivative of $D_{KL}$ has to be non-positive and vanishes at the steady-state:
\begin{equation}
    \mathcal{G} = - \frac{dD_{KL}}{dt}\geq 0.
    \label{Gfunc}
\end{equation}
From explicit calculations of its time derivative:
\begin{eqnarray}
    \mathcal{G}=\dot{S}+\frac{d_t\langle V\rangle}{\sigma}=\dot{S}_{tot}-\dot{S}_{hk}=\dot{S}_{ex}\geq 0
\end{eqnarray}
where we have used Eq.s~\eqref{eq:entropy},\eqref{eq:evolution_potenial},and \eqref{eq:decomp_excess}). Hence, in terms of the geometrical properties of the velocity, the functional reads:
\begin{eqnarray}
    \mathcal{G}&=&\langle \vec{v}*\vec{v}\rangle -\langle \vec{v}*\vec{v}_{\rm st}\rangle\nonumber\\
    &=&\langle \vec{v}*\vec{v}\rangle-\langle \vec{v}*\vec{\nabla}\times \vec{B}\rangle.
\end{eqnarray}

\subsubsection{Trajectory approach}
To gain thermodynamic intuition on the aforementioned functional, we start again from the Langevin description of the system:
\begin{eqnarray}
\dot{x}^{\mu}&=& F^{\mu}+\xi^{\mu}(t),\\
\langle \xi^{\mu}(t)\rangle &=&0,\\
\langle \xi^{\mu}(t)\xi^{\nu}(t')\rangle &=& 2\sigma D^{\mu\nu}\delta(t-t').
\end{eqnarray}
where $\xi^\mu$ is a Gaussian white noise, $F^\mu$ the deterministic force and $D^{\mu\nu}$ is the diffusion matrix. We define the following generic current \cite{pigolotti2017generic,busiello2019hyperaccurate}:
\begin{eqnarray}
\dot{R}&=&\vec{c}_{R}(\vec{x})*\dot{\vec{x}},\\
c_{R}^{\mu}(x)&=&\frac{\partial R}{\partial x^{\nu}} D^{\nu\mu}.
\end{eqnarray}
where $\vec{c}=\vec{c}_{R}(\vec{x})$ is the  vector field that determines the current. By substituting the Langevin dynamics using the Ito formula, we obtain:
\begin{eqnarray}
\dot{R}&=&\vec{c}*\dot{\vec{x}}=\frac{\partial R}{\partial x^{\nu}}\dot{x^{\nu}}+\frac{1}{2}\frac{\partial^2 R}{\partial x^{\mu}\partial x^{\nu}}D^{\mu\nu}\\
&=& \vec{c}*\vec{F}+\sqrt{2} \vec{c}*\vec{\xi}+\vec{\nabla }*(\hat{D} \vec{c})\\
&=&\vec{c}* \vec{v}+\frac{\vec{\nabla}*(\hat{D} \vec{c} P)}{P}+\sqrt{2}\vec{c}*\vec{\xi}.
\end{eqnarray}
Thanks to the explicit equation for the dynamics of the current we can take the average:
\begin{eqnarray}
\langle \dot{R}\rangle = \int d\vec{x} P(\vec{x},t) \int d\vec{\xi} P(\vec{\xi},t) \dot{R} =\langle \vec{c} * \vec{v}\rangle.
\end{eqnarray}
Consider now the following functional along a single stochastic trajectory $\Gamma$:
\begin{eqnarray}
f[\Gamma] = -\log\left(\frac{P(\vec{x},t)}{P_{\rm st}(\vec{x})}\right) \Bigg\vert_{\Gamma}.
\end{eqnarray}
Its average coincides with the Kullback-Leibler divergence and the time derivative reads:
\begin{eqnarray}
g \left[\Gamma\right]&=&\frac{df[\Gamma]}{dt}=-\frac{d\log P}{dt}\Big\vert_{\Gamma}-\dot{x}^{\mu}\partial_{\mu}\log P\vert_{\Gamma}+\dot{x}^{\mu}\partial_{\mu}\log P^*\vert_{\Gamma}\nonumber\\.
\end{eqnarray}
By using the Fokker-Planck equation, we have:
\begin{eqnarray}
g[\Gamma]&=&-\frac{d\log P}{dt}\vert_{\Gamma}-\dot{x}^{\mu}D_{\mu\nu}v^{\nu}\vert_{\Gamma}+\dot{x}^{\mu}D_{\mu\nu}v^*{{\nu}}\vert_{\Gamma}\nonumber\\
&=&-(\vec{\nabla}\cdot \vec{v}-\sigma \vec{v}\cdot\vec{\nabla}\log P)\vert_{\Gamma}+\dot{\vec{x}}* \vec{v}\vert_{\Gamma}\nonumber\\&-&\dot{\vec{x}}*\vec{\nabla}\times \vec{B}\vert_{\Gamma}.
\end{eqnarray}
We can interpret these terms as coming from the variation of the following trajectory-dependent thermodynamic current:
\begin{eqnarray}
g[\Gamma]&=& \dot{s}_{tot}(t)-\dot{s}^{\circ},
\end{eqnarray}
where the first term is the entropy production and the second one is the current surviving at the steady state that depends on the solenoidal part of the force. In fact,
\begin{eqnarray}
\dot{s}_{tot}&=&\dot{s}_{sys}+\dot{s}_{m}=-\partial_{t}\log P\vert_{\Gamma}+\dot{\vec{x}}* \vec{v}\vert_{\Gamma}\\
\dot{s}_{m}&=&\dot{\vec{x}}* \vec{F}=\dot{\vec{x}}* \vec{v}_{\rm st}-\dot{\vec{x}}*\hat{D}\vec{\nabla} V=\dot{s}^{\circ}+\dot{s}^{\Delta}.
\end{eqnarray}
If the Einstein relation $\sigma=k_{B}T$ is valid, the entropy production in the environment can be related to the heat flow in the thermal bath:
\begin{eqnarray}
\dot{s}_{env}=\frac{1}{k_{B}T}\dot{\vec{x}}*\vec{F}=\frac{\dot{q}}{k_{B}T},
\end{eqnarray}
where we have rescaled the diffusive scalar product by the temperature. Following Hatano and Sasa \cite{hatano2001steady} and the seminal work of Sekimoto \cite{sekimoto1998langevin}, one can decompose the heat flow into the excess and housekeeping part:
\begin{eqnarray}
\dot{q}&=&\dot{q}_{ex}+\dot{q}_{hk}=\dot{q}^{\Delta}+\dot{q}^{\circ}\\
\dot{q}_{ex}&=&\dot{q}^{\Delta}=-\dot{\vec{x}}* \hat{D}\vec{\nabla} V\\
\dot{q}_{hk}&=&\dot{q}^{\circ}= \dot{\vec{x}}* \vec{v}_{\rm st}=\dot{\vec{x}}*\vec{\nabla}\times \vec{B}.
\end{eqnarray}
By taking the averages, we obtain:
\begin{eqnarray}
\dot{Q}_{hk}&=&\dot{Q}^{\circ}= \langle \dot{q}_{hk}\rangle =\langle \vec{v}*\vec{\nabla}\times \vec{B}\rangle\\
\dot{S}_{hk}&=&\frac{\dot{Q}_{hk}}{k_{B}T}\\
\dot{Q}_{ex}&=&\dot{Q}^{\Delta}=-\langle v\cdot \nabla V\rangle\\
\dot{S}_{ex}&=&\frac{\dot{Q}_{ex}}{k_{B}T}.
\end{eqnarray}
Hence, the functional is composed by the total entropy production minus the entropy production along cycles in the trajectory space. This last term coincides with the housekeeping heat dissipation in the thermal bath:

\begin{eqnarray}
\mathcal{G}&=&\langle g\rangle_{\Gamma,x} =\langle \dot{s}_{tot}\rangle -\langle \dot{s}^{\circ}\rangle =
\dot{S}_{tot}+\frac{\dot{Q}_{hk}}{k_{B}T}\geq 0
\end{eqnarray}

\section{Multiplicative noise}
\subsection{Geometrical properties of velocities}
Consider a general overdamped system where the diffusion coefficient depend on the variables, i.e. with  multiplicative noise:
\begin{eqnarray}\label{eq:lan_mult}
\dot{x}^{\mu}&=&F^{\mu}(x)+G^{\mu\nu}(x)\xi_{\nu}(t)\\
\langle \xi^{\mu}(t)\rangle &=&0,\\
\langle \xi^{\mu}(t)\xi^{\nu}(t')\rangle &=& 2C^{\mu\nu}\delta(t-t').
\end{eqnarray}
with $\hat{G}$ the matrix ruling the multiplicative fluctuations, that for simplicity we consider diagonal $\hat{G}(\vec{x}) = {\rm diag}(g_{1}(\vec{x}), g_2(\vec{x}), \dots, g_{N}(\vec{x})$, and $\hat{C}$ the correlation matrix of the Gaussian noise ($\vec{x}$-independent). In the following, we employ an $\alpha$-dependent discretization of the noise in order to encode all possible prescription. Indeed, $\alpha=0$ corresponds to the Stratonovich case, while $\alpha = 1$ to the Ito prescription. The corresponding Fokker-Planck equation reads:
\begin{eqnarray}
    \partial_{t}P&=&-\partial_{\mu}J^{\mu}\\
    J^{\mu}&=&F^{\mu}P-\alpha\partial_{\nu}D^{\mu\nu}-D^{\mu\nu}\partial_{\nu}P,
\end{eqnarray}
where the diffusion matrix is constructed as follows:
\begin{eqnarray}\label{eq:trans_diff}
\hat{D} = \hat{G}^{T}\hat{C}\hat{G}.
\end{eqnarray}
In this general case, Eq.~\eqref{eq:potential} is not valid anymore and the stationary solution in the small noise limit is:
\begin{eqnarray}
P_{\rm st}(\vec{x})=\eta(\sigma)z(\vec{x})e^{-\frac{V}{\sigma}}.
\end{eqnarray}
The stationary velocity is defined now as:
\begin{eqnarray}
v^{\mu}_{\rm st} &=& F^{\mu} + D^{\mu\nu} \partial_{\nu} V + \nonumber\\
&-& \alpha \sigma \partial_{\nu} D^{\mu\nu} - \sigma D^{\mu\nu} \partial_{\nu} \log z \nonumber\\
&=& (v^{(0)})^{\mu}_{\rm st}+\sigma (v^{(1)})^{\mu}_{\rm st}.\\
(v^{(0)})^{\mu}_{\rm st}&=&F^{\mu}+D^{\mu\nu}\partial_{\nu}V\\
(v^{(1)})^{\mu}_{\rm st}&=&-\alpha\partial_{\nu} D^{\mu\nu}- D^{\mu\nu}\partial_{\nu}\log z.
\end{eqnarray}
We solve order by order in $\sigma$ the stationary Fokker-Planck equation. The dominant $\sigma^{-1}$ term does not change with respect to the additive case:
\begin{eqnarray}\label{eq:zeroth_v}
\left(F^{\mu}+D^{\mu\nu}\partial_{\nu}V\right)\partial_{\mu}V&=&0\nonumber\\
\vec{v}^{(0)}_{\rm st}\cdot\vec{\nabla} V=0.
\end{eqnarray}
This condition, as in the constant case, tells us that the first order velocity is tangent to the height lines of the non-equilibrium potential. Assuming that the potential has just one minimum, by using the Gauss' law, we can show this geometrical property of $\vec{v}^{(0)}$ coincides the fact that it also has zero divergence. Consider the volume $\Omega$ enclosed by a single level surface of the potential $V$. The vector pointing outward in each point of the boundary is the gradient of the potential. From the Gauss' law:
\begin{eqnarray}
    \int_{\Omega}\vec{\nabla}\cdot \vec{v}^{(0)}_{\rm st}d\Omega=-\int_{\Sigma=\partial\Omega}d\Sigma~\vec{v}^{(0)}_{\rm st}\cdot \frac{\vec{\nabla}V}{\vert \vec{\nabla}V\vert}=0.
\end{eqnarray}
Given that for any point in the phase space there exists a contour line passing through it, the divergence is always zero, i.e. $\vec{\nabla} \cdot \vec{v}^{(0)}_{\rm st}=0$.

The $\sigma^0$ order has two new terms with respect to the additive scenario:
\begin{eqnarray}\label{eq:first_v}
&&(F^{\mu}+D^{\mu\nu}\partial_{\nu}V)\partial_{\mu}\log z+\partial_{\mu}\left(F^{\mu}+\frac{D^{\mu\nu}}{2}\partial_{\nu}V\right)\nonumber\\&+&\left(D_{\mu\nu}\partial^{\nu}\log z+\alpha\partial^{\nu}D_{\mu\nu}\right)\partial^{\mu} V=0\\
&=&\vec{\nabla}\cdot \vec{v}^{(0)}_{\rm st}+\vec{v}^{(0)}_{\rm st}\cdot\vec{\nabla} \log z-\vec{v}^{(1)}_{\rm st}\cdot\vec{\nabla} V=0
\end{eqnarray}
Finally, the  order $\sigma$ reads:
\begin{eqnarray}\label{eq:second_v}
&&\partial_{\mu}\left(\alpha\partial_{\nu}D^{\mu\nu}+D^{\mu\nu}\partial_{\nu}\log z\right)+\nonumber\\
&+&\left(D^{\mu\nu}\partial_{\nu}\log z+\alpha\partial_{\nu}D^{\mu\nu}\right)\partial_{\nu}\log z\nonumber\\
&=&\vec{\nabla}\cdot \vec{v}^{(1)}_{\rm st}+\vec{v}^{(1)}_{\rm st}\cdot\vec{\nabla} \log z =0.
\end{eqnarray}
To go further and investigate the geometric properties of the first order velocity, let's recall that by a simple change of variable it is possible to transform the multiplicative noise of the Langevin equation into an additive one. If $\hat{\Lambda}$ is the Jacobian of the transformation, i.e., $\vec{x'}=\hat{\Lambda}\vec{x}$, this mapping is done by requiring that $\hat{\Lambda}=\hat{G}^{-1}$. In these new coordinates, all the results obtained before straightforwardly hold:
\begin{eqnarray}
    P'=\frac{e^{-V(x')/\sigma}}{Z}.
\end{eqnarray}
Hence, by writing down the transformation of $P$, we have:
\begin{eqnarray}\label{eq:change_p}
P&=&P'\big\vert \hat{\Lambda}\big\vert\nonumber\\
P&=&\frac{1}{Z}\vert \hat{\Lambda}\vert e^{-V(x'(x))/\sigma}
\end{eqnarray}
where $\vert \cdot\vert$ is the determinant. Hence, we identify:
\begin{eqnarray}
    z=\vert \hat{\Lambda}\vert =\vert \hat{G}\vert ^{-1} .
\end{eqnarray}
As a consequence, the first order velocity is proportional to the gradient of the transformation plus a $\alpha$-dependent term coming from the change of coordinates:
\begin{eqnarray}
    (v^{(1)}_{\rm st})^\mu &=& -\alpha\partial_{\nu}D^{\mu\nu}-D^{\mu\nu}\partial_{\nu}\log\vert\Lambda\vert\\
   &=&-\alpha\partial_{\nu}D^{\mu\nu}+D^{\mu\nu}\partial_{\nu}\log\vert G\vert\nonumber\\
   &=&\alpha\partial_{\nu}D^{\mu\nu}+D^{\mu\nu}\partial_{\nu}\psi
\end{eqnarray}
Putting all these ingredients together, the functional can be written as follows:
\begin{eqnarray}
    \mathcal{G}&=&\langle \vec{v}*\vec{v}\rangle-\langle \vec{v}*\vec{v}_{\rm st}\rangle\\
    &=&\langle \vec{v}*\vec{v}\rangle-\langle \vec{v}*\vec{\nabla}\times \vec{B}\rangle-\sigma\langle v*D\vec{\nabla}\psi\rangle-\alpha\sigma\langle \vec{v}*\vec{\nabla} D\rangle.\nonumber
\end{eqnarray}

\section{Examples}
\subsection{Driven brownian particle on a torus}
Consider a Brownian particle confined in a torus by a quadratic potential, $U(x,y)$, and driven along the torus itself by a constant driving force, $f$. Here, $f$ breaks the detailed balance, and the stationary polar flux will induce a topological symmetry breaking in the system. Let us start with the description of the Brownian motion:
\begin{eqnarray}
\dot{x} = f_x(x,y) - \partial_x U(x,y,z) + \sqrt{2 D} ~\xi_x(t) \nonumber \\
\dot{y} = f_y(x,y) - \partial_y U(x,y,z) + \sqrt{2 D} ~\xi_y(t) \\
\dot{z} = f_z(x,y) - \partial_z U(x,y,z) + \sqrt{2 D} ~\xi_z(t) \nonumber
\end{eqnarray}
where $f_i$ is the component of the driving force along $i$ and $\xi_i(t)$ is a Gaussian white noise with unit variance. These equations can be rewritten in polar coordinates $(\rho,\phi,z)$, with $\rho = \sqrt{x^2 + y^2}$ and $\phi = \arctan{(y/x)}$. Employing the Ito's formula for the change of variable, we obtain:
\begin{eqnarray}
\dot{\rho} &=& - \gamma (\rho - \rho^*) + \frac{D}{\rho} + \sqrt{2 D} ~\xi_\rho (t) \nonumber \\
\dot{\phi} &=& \frac{f}{\rho} + \frac{\sqrt{2 D}}{\rho} \xi_\phi (t) \\
\dot{z} &=& - \gamma z + \sqrt{2 D} ~\xi_z (t) \nonumber
\end{eqnarray}
with $U = (\gamma/2) (\rho - \rho^*)^2 + (\gamma/2) z^2$, $f$ only acts along $\phi$, and $D/\rho$ is the usual Ito term. The resulting Fokker-Planck equation is:
\begin{equation}
    \partial_t P = (\mathcal{L}_\rho + \mathcal{L}_\phi + \mathcal{L}_z) P
\end{equation}
where we defined:
\begin{eqnarray}
\mathcal{L}_\rho &=& \partial_\rho \left( \gamma (\rho - \rho^*) - \frac{D}{\rho} \right) + D ~\partial_{\rho\rho} \nonumber \\
\mathcal{L}_\phi &=& \partial_\phi \left( -\frac{f}{\rho} \right) + \frac{D}{\rho^2} ~\partial_{\phi\phi} \\
\mathcal{L}_z &=& \partial_z \left( \gamma z \right) + D ~\partial_{zz} \nonumber
\end{eqnarray}
The motion along $z$ is decoupled, while the other two equations can be solved imposing that the flux only flows along $\phi$. The steady-state is:
\begin{equation}
    P_{\rm st} = Z^{-1} \exp\left( - \frac{\rho (\rho - 2 \rho^*) \gamma + \gamma z^2 - 2 D \log \rho}{2 D}\right)
\end{equation}
Hence, in the small noise limit, we have:
\begin{equation}
    P_{\rm st} \simeq Z^{-1} \exp\left( - \frac{\rho (\rho - 2 \rho^*) \gamma + \gamma z^2}{2 D}\right)
\end{equation}
which is of the form outlined in the main text. The flux only acts along $\phi$, and it is equal to:
\begin{equation}
    \vec{J}_{\rm st} = Z^{-1} \exp\left( - \frac{\rho (\rho - 2 \rho^*) \gamma + \gamma z^2}{2 D}\right) f ~\boldsymbol{i}_\phi
\end{equation}
where $\boldsymbol{i}_\phi$ is the versor indicating the coordinate $\phi$. Hence, the stationary velocity reads:
\begin{equation}
    \vec{v}_{\rm st} = - \frac{f}{\rho} ~\boldsymbol{i}_\phi
\end{equation}

\subsection{Brownian Gyrator}
Consider the two dimensional motion in the $x-y$ plane of a particle under the effect of a parabolic potential $U$ and two baths at temperatures $T_{x}=T$ and $T_{y}=T(1+\delta)$, each one acting along a different direction:
\begin{eqnarray}
\dot{x}&=&-\partial_{x}U(x,y)+\xi_{x}(t)\nonumber\\
\dot{y}&=&-\partial_{y}U(x,y)+\xi_{y}(t)\nonumber\\
\langle \xi_{i}(t)\rangle&=&0\nonumber\\
\langle \xi_{i}(t)\xi_{j}(t')\rangle&=&2 T_{i}\delta_{ij}\delta(t-t'),
\end{eqnarray}
where the potential is $U(x,y)=\frac{x^2}{2}+\frac{y^2}{2}+uxy$, with $\vert u\vert <1$ to confine the particle near the origin.
The system approaches the following stationary solution \cite{cerasoli2018asymmetry}:
\begin{eqnarray}
P_{\rm st}(x,y)&=&Z^{-1}e^{-\frac{V}{T\eta}}\\
V(x,y)&=&\frac{\gamma_{1}}{2}x^2+\frac{\gamma_{2}}{2}y^2+u\gamma_{3}xy,
\end{eqnarray}
where we introduced
\begin{eqnarray*}
\eta&=&1+\delta+\frac{u^2}{4}\delta^2\\
\gamma_{1}&=&1+\delta- \frac{u^2}{2}\delta\\
\gamma_{2}&=&1+ \frac{u^2}{2}\delta\\
\gamma_{3}&=&(2+\delta)
\end{eqnarray*}
In the main text we have renamed $\sigma=T\eta$ to simplify the exposition.
Note that the stationary solution respects the non-equilibrium potential ansatz, where $T$ is the parameter regulating the noise, while $\delta$ and $u$ affect the detailed balance condition that reads as follows:
\begin{eqnarray}
T_{x}\partial_{x}F_{y}=T_{y}\partial_{y}F_{x}\rightarrow u\delta =0.
\end{eqnarray}
Due to the anisotropy of the temperatures, detailed balance is broken and a current emerges with a velocity:
\begin{eqnarray}
v_{x}&=&\partial_{x}U-\eta^{-1}\partial_{x}V\\
v_{y}&=&\partial_{y}U-\eta^{-1}(1+\delta)\partial_{y}V.
\end{eqnarray}
Its divergence reads:
\begin{eqnarray}
\vec{\nabla}\cdot \vec{v}_{\rm st}=\frac{1}{\eta}\left(2\eta -\gamma_{1}-(1+\delta)\gamma_{2}\right)=0,
\end{eqnarray}
thus the vorticity is:
\begin{eqnarray}
w&=&\partial_{x}v_{y}-\partial_{y}v_{x}=\frac{u\gamma_{3}(T_{y}-T_{x})}{\eta}=-\frac{u\delta\gamma_{3}}{\eta}.
\end{eqnarray}
In terms of just the temperatures and the potential tilting the vorticity  reads:
\begin{equation}
    w=\frac{4 T_x u}{4 T_x T_y + (T_x - T_y)^2 u^2} \left( T_x^2 - T_y^2 \right)
\end{equation}
Hence, the interplay between particle interaction and temperature difference is accountable for the emergence of a dissipative symmetry breaking in the NESS.


\end{document}